\documentclass{elsart}
\usepackage{amssymb,epsfig}

\begin{document}
\begin{frontmatter}

\title{Anomalous diffusion in non-equilibrium relativistic heavy-ion 
rapidity spectra}
\vspace{-1cm}
\author{A. Lavagno}
\address{
Dipartimento di Fisica, Politecnico di Torino and INFN, 
Sezione di Torino, Torino, Italy}
\date{\today }
\maketitle

\begin{abstract}      
There is experimental and theoretical evidence that the broad rapidity 
distribution of net proton yield ($p-\overline{p}$) in central 
heavy-ion collisions at SPS energies could be a signal of  
non-equilibrium properties of the system. 
We show that the broad rapidity shape can be well reproduced in the framework 
of a non-linear relativistic 
Fokker-Planck dynamics which incorporates non-extensive 
statistics and anomalous diffusion. 
\end{abstract}
\end{frontmatter}
\vspace{-0.5cm}

\vspace{-0.5cm}
\section{Introduction}
\vspace{-0.5cm}
In relativistic heavy-ion collisions nuclear matter 
is believed to be compressed 
to high baryon density and nuclei undergoing central collisions 
reduce strongly 
their original longitudinal momentum. This loss of rapidity, 
usually referred as baryon stopping, is an important characteristic 
to understand the reaction mechanism at high energy density. 


Isotropic thermal models fail to reproduce the broad and flat 
rapidity distribution observed in the central Pb-Pb collisions 
at $158$ A GeV/c \cite{na49}. 
Collective and anisotropic flows must be considered 
(with the introduction of new parameters) to take into account 
the experimental spectra \cite{na44,braun}. 
Complicate evolution codes take only partially into account of such 
distributions and the physical evolution of the system is not 
completely understood \cite{cape}.


In this paper we study the evolution of the rapidity distribution from 
a macroscopic point of view by using  
a non-linear relativistic Fokker-Planck equation and we 
show that the observed broad rapidity shape could be a signal of 
non-equilibrium properties of the system. 
Let us note that a similar approach, within a linear Fokker-Planck equation, 
has been previously studied in Ref.\cite{wol} by using a linear drift in the 
space of the rapidity and a free parameter diffusion coefficient. With this 
choice, the author found a strongly violation of the fluctuation-dissipation 
theorem. We will see that generalizing the Brownian motion to the relativistic 
kinetic variables, the standard Einstein relation is satisfied and 
Tsallis non-extensive statistics emerges in a natural way from the 
non-linearity of the Fokker-Planck equation.

\vspace{-0.5cm}
\section{Non-linear Fokker-Planck equation in relativistic dynamics}
\vspace{-0.5cm}
A class of anomalous diffusions are currently described through the 
non-linear Fokker-Planck equation (NLFPE)
\begin{equation}
\frac{\partial}{\partial t}[f(y,t)]^{\mu}=\frac{\partial}{\partial y} 
\left [ J(y) [f(y,t)]^\mu+D \frac{\partial}{\partial y}[f(t,y)]^\nu
\right ] \; ,
\label{nlfpe}
\end{equation}
where $D$ and $J$ are the diffusion and drift coefficients, respectively. 
Tsallis and Bukman \cite{tsallis} have shown that, for 
linear drift, the time dependence solution of the above equation is a 
Tsallis-like distribution with $q=1+\mu-\nu$. The norm of the distribution is 
conserved for all times only if $\mu=1$, therefore we will 
limit the discussion to the case $\nu=2-q$.

We are going to study the time evolution of the particle distribution  
by means the above NLFPE in the rapidity space. 
Basic assumption of our analysis is that the rapidity distribution $y$ is 
not appreciably influenced by transverse dynamics which is considered 
in thermal equilibrium. Such hypothesis is well 
confirmed by the experimental data and adopted in many theoretical 
works \cite{na49,na44,braun}. 
A crucial r\^ole in the solution of the above NLFPE 
plays the choice of the diffusion and the drift coefficients. 
Such a choice influences the time evolution of the system  and its equilibrium 
distribution. 

Since the temperature at freeze-out exceeds $100$ MeV, Boltzmann 
approximation is usually adopted and the single particle equilibrium 
distribution is written as
\begin{eqnarray}
E\frac{d^3N}{d^3p}\propto E\, \exp(-E/T) 
\equiv m_\perp \, \cosh(y) \, 
\exp(-m_\perp \cosh(y)/T) \; ,
\label{distri}
\end{eqnarray} 
where $y$ is the rapidity, $m_\perp=\sqrt{m^2+p_\perp^2}$ 
is the transverse mass, $T$ is the temperature. 
Imposing the validity of the Einstein 
relation for Brownian particle, we can generalize to the relativistic case 
the standard expressions of diffusion and drift coefficients as follows 
\begin{equation}
D=\gamma \, T\; , \ \ \ \  \ J(y)=\gamma \, m_\perp \sinh(y)
\equiv  \gamma \, p_\parallel \; ,
\label{coeffi}
\end{equation}
where $p_\parallel$ is the longitudinal momentum and $\tau$ 
is a common constant. 
It is easy to see that the above coefficients give 
us the Boltzmann stationary distribution (\ref{distri}) in the 
linear Fokker-Planck equation ($q=\nu=1$)\footnote{This result cannot be 
obtained if one assumes a linear drift 
coefficient as in Ref.\cite{wol}.}, while  
the stationary solution of the NLFPE (\ref{nlfpe}) 
with $\nu=2-q$ is a Tsallis-like 
distribution with the relativistic energy $E=m_\perp \cosh(y)$. 
In the light of the above results, we generalize the results of Tsallis-Bukman  
by searching as a solution of the NLFPE (\ref{nlfpe}) the following 
time dependent Tsallis-like distribution 
\begin{equation}
f_q(y,m_\perp,t)=
\Big \{1-(1-q)\,\beta(t)\, m_\perp \cosh[y-y_m(t)]\Big\}^{1/(1-q)} \; .
\end{equation}
The unknown functions $\beta (t)=1/T(t)$ and $y_m(t)$ 
have been derived by means numerical integration of 
Eq.(\ref{nlfpe}) with initial $\delta$-function condition depending 
on the value of the experimental projectile rapidities.  
The rapidity distribution at fixed time is then 
obtained by numerical integration over the transverse mass 
$m_\perp$ (or transverse momentum) as follows
\begin{equation}
\frac{dN}{dy}(y,t)=c\, \int_m^\infty \!\! m^2_\perp \, \cosh(y) 
\, f_q(y,m_\perp,t) \, dm_\perp \; ,
\label{raint}
\end{equation}
where $c$ is the normalization constant fixed by the total number of the 
particles. 
Rapidity spectra calculated from (\ref{raint}) will ultimately depend on two 
parameters: the ``interaction'' time $\tau=\gamma t$ and the non-extensive 
parameter $q$. Therefore, no more free parameters are used 
in this analysis respect to previous theoretical studies \cite{na44,braun,cape}.

\vspace{-0.5cm}
\section{Results and Conclusions}
\vspace{-0.5cm}
We have studied the time dependent rapidity distribution (\ref{raint}) 
in the case of central  
Pb+Pb collisions at 158 GeV/c and our results are compared with the 
net proton experimental NA49 data \cite{na49}. 
The obtained spectra are normalized 
to 164 protons and the beam rapidity is fixed to $y_{\rm cm}=2.9$ (in the 
c.m. frame) \cite{na49}. 
In Fig.1 we show the calculated rapidity spectra compared with the experimental 
data. The full line corresponds to the NLFPE solution (\ref{raint}) at 
$\tau=0.82$ and $q=1.25$; the dashed line corresponds to the solution 
of the linear case ($q=1$) at $\tau=1.2$. 
Only in the non-linear case ($q\ne 1$) exists a (finite) time for which 
the obtained rapidity spectra well reproduces the broad experimental shape. 
A value of $q\ne 1$ implies anomalous superdiffussion in the rapidity space, 
i.e., $[y(t)-y_M(t)]^2$ scale like $t^\alpha$ with $\alpha>1$ \cite{tsallis}.

\begin{figure}[ht]
\parbox{6cm}{
\scalebox{0.6}{
\includegraphics*[-10,450][570,770]{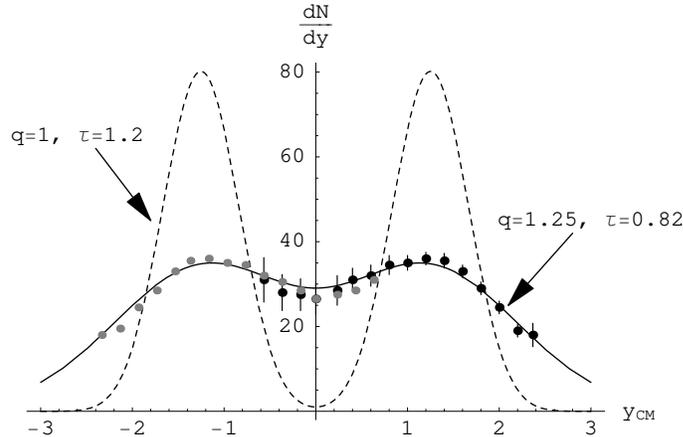}}
}

\parbox{14cm}{\vspace{-1cm}
\caption{
Rapidity spectra for net proton production ($p-\overline{p}$) in 
central Pb+Pb collisions at 158A GeV/c (grey circles are data reflected about 
$y_{cm}=0$) \cite{na49}. Full line corresponds 
to our results by using a non-linear evolution equation ($q=1.25$), 
dashed line corresponds to the linear case ($q=1$). 
}}
\vspace{0.5cm}
\end{figure}

We have studied the rapidity spectra from a 
macroscopic point of view. A good agreement with the experimental 
data is reached assuming a deviation from the equilibrium and a non-linear 
evolution which implies non-extensive Tsallis-like distribution. 
Non-extensive features in relativistic collisions are outlined in 
several works \cite{lava,wilk,rafe,bedia,beck}, 
even if a microscopic justification of these effects is still lacking. 
In Ref.\cite{cape}, the authors show that diquark breaking during inelastic 
collisions in dual parton model could be responsible for the observed baryon 
stopping. The possible connection between the nature of these inelastic collisions
and the above observed non-extensive statistical effects is under investigations.


\noindent
{\it Acknowledgments.} It is a pleasure to thank Prof. P. Quarati and Prof. 
G. Wilk for fruitful discussions and comments.


\vspace{-0.5cm}
\begin{thebibliography}{30}
\vspace{-0.5cm}
\bibitem{na49}
H. Appelsh\"auser et al. (NA49 Collaboration), Phys. Rev. Lett. 82 (1999) 2471.
\bibitem{na44}
I.G. Bearden et al. (NA44 Collaboration), Phys. Rev. Lett. 78 (1997) 2080.
\bibitem{braun}
P. Braun-Munzinger et al., Phys. Lett. B 344 (1995) 43; B 356 (1996) 1.
\bibitem{cape}
A. Capella, C.A. Salgado, Phys. Rev. C 60 (1999) 054906.
\bibitem{wol}
G. Wolschin, Eur. Phys. J. A 5 (1999) 85.
\bibitem{tsallis}
C. Tsallis, D.J. Bukman, Phys. Rev. E 54 (1996) R2197. 
\bibitem{lava}
W.M. Alberico, A. Lavagno, P. Quarati, Eur. Phys. J. (2000) 499.
\bibitem{wilk}
G. Wilk, Z. Wlodarczyk, Phys. Rev. Lett. 84 (2000) 2770.
\bibitem{rafe}
D.B. Walton, J. Rafelski, Phys. Rev. Lett. 84 (2000) 31.
\bibitem{bedia}
I. Bediaga, E.M.F. Curado, J.M. de Miranda, Physica A 286 (2000) 156.
\bibitem{beck}
C. Beck, Physica A 286 (2000) 164.

\end{thebibliography}
\end{document}